\documentclass[twocolumn,showpacs,preprintnumbers,amsmath,amssymb,prb]{revtex4}
\usepackage[dvips]{color,graphics,epsfig,rotating}
\usepackage{graphicx}
\usepackage{dcolumn}
\usepackage{bm}

\begin{document}

\title{Electronic Structure and Doping in BaFe$_2$As$_2$ and LiFeAs:
Density Functional Calculations}

\author{D.J. Singh}
\affiliation{Materials Science and Technology Division,
Oak Ridge National Laboratory, Oak Ridge, Tennessee 37831-6114} 

\date{\today} 

\begin{abstract}
We report density functional calculations of the electronic structure
and Fermi surface of the BaFe$_2$As$_2$
and LiFeAs phases including doping via the virtual crystal approximation.
The results show that contrary to a rigid band picture, the density of
states at the Fermi energy is only weakly doping dependent and that
the main effect of doping is a change in the relative sizes of the
electron and hole Fermi surfaces as required by Luttinger's theory.
This is a consequence of a change in As height with doping, in particular
a shift of As towards Fe as holes are introduced in the Fe plane, as might
be expected from simple ionic considerations.
The main effect of doping is therefore a reduction in the degree of nesting
of the Fermi surface. This provides a framework for understanding
the approximate electron-hole symmetry in the phase diagrams of the
Fe-As based superconductors.
\end{abstract}


\maketitle

\section{introduction}

The discovery of high temperature superconductivity in oxypnictide
phases, prototype LaFeAs(O,F) by Kamihara and co-workers \cite{kamihara}
has sparked widespread interest in establishing the physical properties
of these materials and especially the mechanism for superconductivity.
This interest has led to the discovery of a number of new phases
including other oxy-arsenides that when electron
doped have critical temperatures exceeding 55K, \cite{cwang}
and hole doped superconducting phases including (K,Ba)Fe$_2$As$_2$,
with $T_c$=38K \cite{rotter1,rotter2} and Li$_{1-y}$FeAs with $T_c$
of 18K. \cite{xcwang}
Thus there are three families discovered so far: (1)
the oxy-arsenides, spacegroup $P4/nmms$, with rare earth oxide layers
separating the FeAs layers common to all of these superconductors,
(2) the body centered tetragonal, $I4/mmm$
ThCr$_2$Si$_2$ structure materials (BaFe$_2$As$_2$),
where the FeAs layers are stacked so that the As
atoms face each other and the Ba atoms sit in the resulting 8-fold
coordinated square prismatic sites between them,
and (3) $P4/nmms$ LiFeAs, which is like the oxy-arsenide structure but
with the rare earth oxide layers removed
and replaced by Li (see below), resulting in a large
decrease in the $c$-axis spacing of the FeAs layers.
Stoichiometric
(undoped) BaFe$_2$As$_2$ shows a spin density wave (SDW) type 
magnetic order and a structural transition with similar ordering
temperatures.
This feature is common to most of these materials, including the
oxy-arsenides. \cite{cruz,huang}
However, there are also some strong differences between properties
of the oxypnictides and BaFe$_2$As$_2$. Most notable is the fact that
with the exception of one report,
\cite{wen-epl}
the oxy-arsenide superconductors are generally electron doped,
while (Ba,K)Fe$_2$As$_2$ and Li$_{1-y}$FeAs
are found to superconduct when hole doped.
However, in both BaFe$_2$As$_2$ and LaFeAsO superconductivity
apparently is associated with the suppression of the SDW.
This is potentially of importance because there has been
much discussion of the association between magnetic quantum
critical points and high temperature superconductivity
and in addition the specific spin fluctuations associated with the
nesting that presumably drives the SDW have been studied as
a possible pairing mechanism. \cite{mazin}

Electronic structure studies,
\cite{singh-du,nekrasov-basr,ma-ba,nekrasov-li}
show that the Fermi surfaces and band structures of these phases are
qualitatively rather similar.
In particular, all the compounds show small compensating electron
and hole Fermi surfaces, but high densities of states.
The hole Fermi surfaces occur around the zone center, and are generally
derived from heavier (lower velocity) bands than the electron surfaces,
which are around the zone corner ($M$ point in a primitive tetragonal zone).
Within band structure theories, moment formation is governed by a Stoner
parameter, $I$, which takes values of 0.7 eV to 0.9 eV for ions near the
middle of the $3d$ series (note that the effective $I$ can reduced by
hybridization). While magnetism may occur with lower values of the density
of states, it must occur within a band picture if the Stoner criterion,
$N(E_F)I > 1$, is met (the non-spin-polarized electronic
structure becomes unstable against ferromagnetism in this case).
Here $N(E_F)$ is the
density of states (DOS) at the Fermi energy on a per atom per spin basis.
The calculated values of $N(E_F)$ for the undoped FeAs materials put
them at the borderline of itinerant magnetism, and in fact they are
near both ferromagnetism, as required by the Stoner theory, and checkerboard
antiferromagnetism, where nearest neighbor Fe atoms are oppositely polarized.
The ground states, both experimentally \cite{cruz,huang,goldman}
and theoretically \cite{mazin,ma-ba,dong,ma,yildirim,yin}
have a magnetic
structure that corresponds to condensation of an $M$ point (zone corner) SDW.

The electronic structures all show a strong increase in the electronic DOS
as the Fermi energy is lowered into the heavier hole bands.
This feature is quite
robust in this family and is also the case for the corresponding phosphide
compounds.
\cite{lebegue,gustenau}
Importantly, this strong increase in the DOS poses a conundrum because it
means that within a rigid band picture hole doped materials would be strongly
magnetic, and therefore would necessarily have very different properties,
including superconductivity, from the electron doped materials.
On the other hand, from an experimental point of view the
generic phase diagram is much more symmetric, with an SDW and structural
distortion near zero doping, with apparently non-magnetic, and superconducting
states for both hole and electron doped materials.
Here we investigate this issue using calculations for doped
BaFe$_2$As$_2$ and LiFeAs within the virtual crystal approximation.
This is an average potential approximation, which is
beyond the rigid band method. We in fact find strong non-rigid band
effects that leave only a weak doping level dependence to the $N(E_F)$
and provide a
plausible explanation for the generic phase diagram of this family of
materials.

\section{structure and method}

The present
calculations were performed within the
local density approximation (LDA) using the general potential linearized
augmented planewave (LAPW) method, \cite{singh-book}
similar to those reported previously for LaFeAsO. \cite{singh-du}
LAPW sphere radii of 1.8 $a_0$, 2.2 $a_0$, 2.1 $a_0$, and 2.1 $a_0$
were used for
Li, Ba, Fe and As, respectively.
The effect of doping was included by the virtual crystal approximation,
by varying the nuclear charge of Ba or Li. Importantly,
the internal parameters, including the As and Li positions,
were relaxed independently for each doping level.

\begin{figure}
\includegraphics[width=3.2in,angle=0]{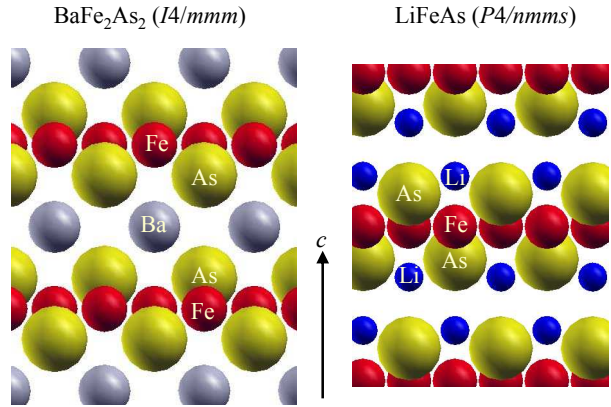}
\caption{\label{struct} (Color online)
Crystal structure of BaFe$_2$As$_2$ (left)
and LiFeAs (right) with the relaxed As and Li coordinates.
Note that in both structures the FeAs layers are uniformly spaced
along the $c$-axis direction.
}
\end{figure}

We took the experimental tetragonal lattice parameters
of room temperature BaFe$_2$As$_2$ of Ref. \onlinecite{rotter1},
and relaxed the internal coordinate of As using LDA total energy
minimization. The calculated coordinate is $z_{\rm As}$=0.342.
This is noticeably lower than the reported value of $z_{\rm As}$=0.3545.
In fact the As heights
in these two structures differ by 0.16 \AA, which
as discussed below is outside the normal range of density functional
errors and is
significant for the electronic structures.
We also did calculations with the generalized gradient approximation
but find a similar discrepancy.
Similar discrepancies have also been noted in the LaFeAsO series. 
\cite{yin,mazin2}
For LiFeAs, we used the experimental lattice parameters\cite{juza}
and relaxed the free internal coordinates.
The experimental structure (spacegroup $P4/nmms$, 129) shows
two Li sites: 2$c$, which according to Ref. \onlinecite{juza}
is fully occupied, and lies above and
below the centers of the Fe squares opposite the As, as shown in
Fig. \ref{struct}, and 2$b$ which has a reported occupancy of 0.1
in Li$_{1.1}$FeAs and lies directly above the Fe.
We performed total energy calculations for stoichiometric
LiFeAs with Li in the 2$b$ and 2$c$ sites and found that the
$2c$ site is favored by 0.52 eV per formula unit.
Accordingly, we performed our doping studies placing Li in the $2c$
site.
We note that while this is the opposite choice to that made by
Nekrasov and co-workers \cite{nekrasov-li}, the electronic
structure near the Fermi energy is practically unaffected by this choice.
In fact, when hole doped by Li vacancies one may expect Coulomb repulsions
between the like charged Li ions to favor partial occupations of the $2b$
site, though always with a higher occupancy on the $2c$ site, similar to
what is found in Na$_x$CoO$_2$.
In any case, the insensitivity of the electronic structure to the Li
position supports the validity of the virtual crystal approximation.
The structures that we used for the undoped compounds are summarized
in Table \ref{tab-struct}.

\begin{table}[tbp]
\caption{Structures used in calculations for undoped BaFe$_2$As$_2$
and LiFeAs.}
\label{tab-struct}
\begin{tabular}{lcc}
\hline
      &  BaFe$_2$As$_2$ & LiFeAs \\
\hline
Spacegroup    &  $I4/mmm$  &  $P4/nmms$ \\
$a$(\AA)      &   3.9625   &  3.776 \\
$c$(\AA)      &   13.0168   &  6.349 \\
$z_{\rm As}$  &   0.3545    &  0.2116 \\
$z_{\rm Li}$  &       &  0.6728 \\
$d$(As-Fe)(\AA) & 2.315    &   2.403 \\
\hline
\end{tabular}
\end{table}

\section{undoped reference systems}

\begin{figure}
\includegraphics[height=3.3in,angle=270]{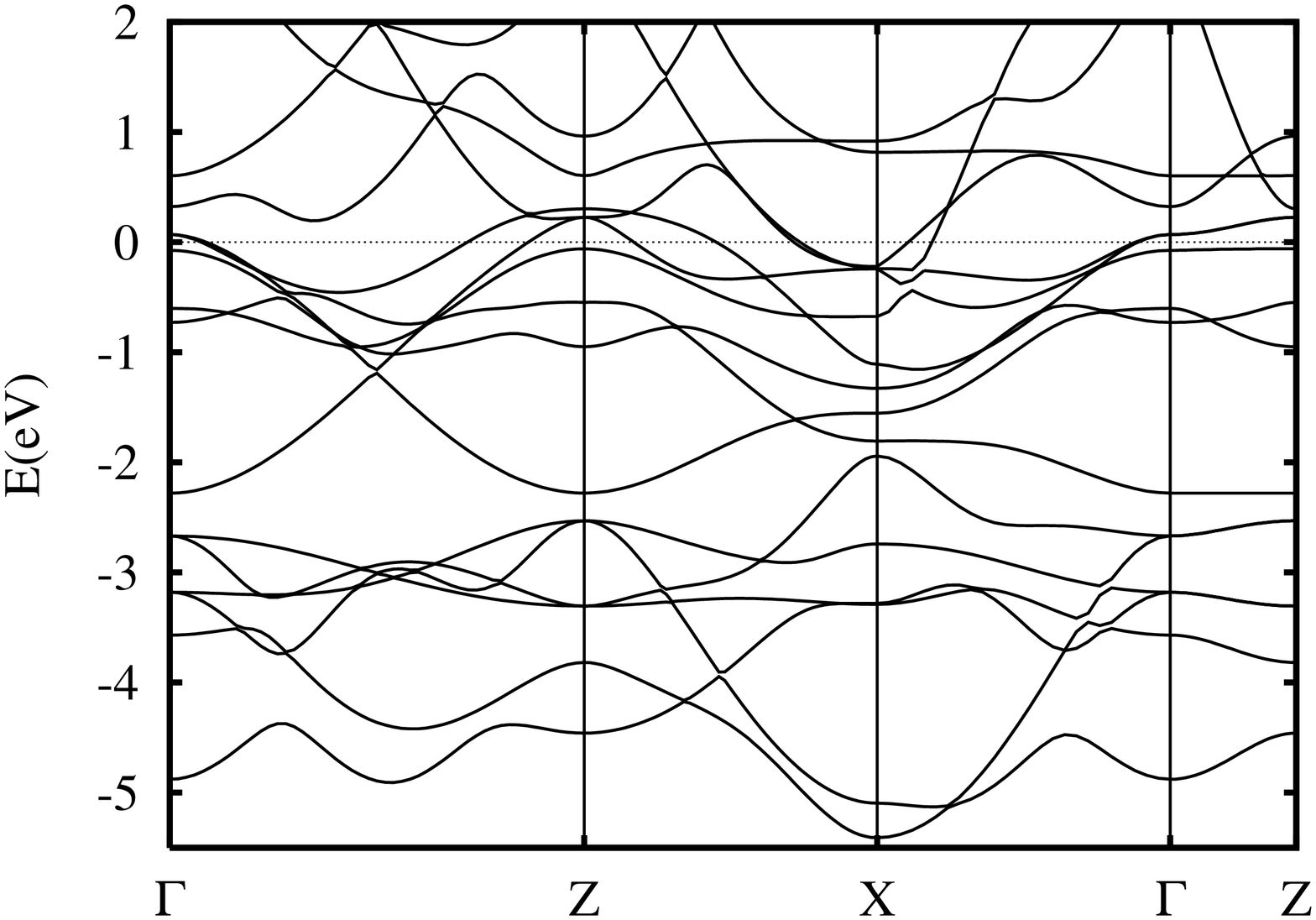}
\includegraphics[height=3.3in,angle=270]{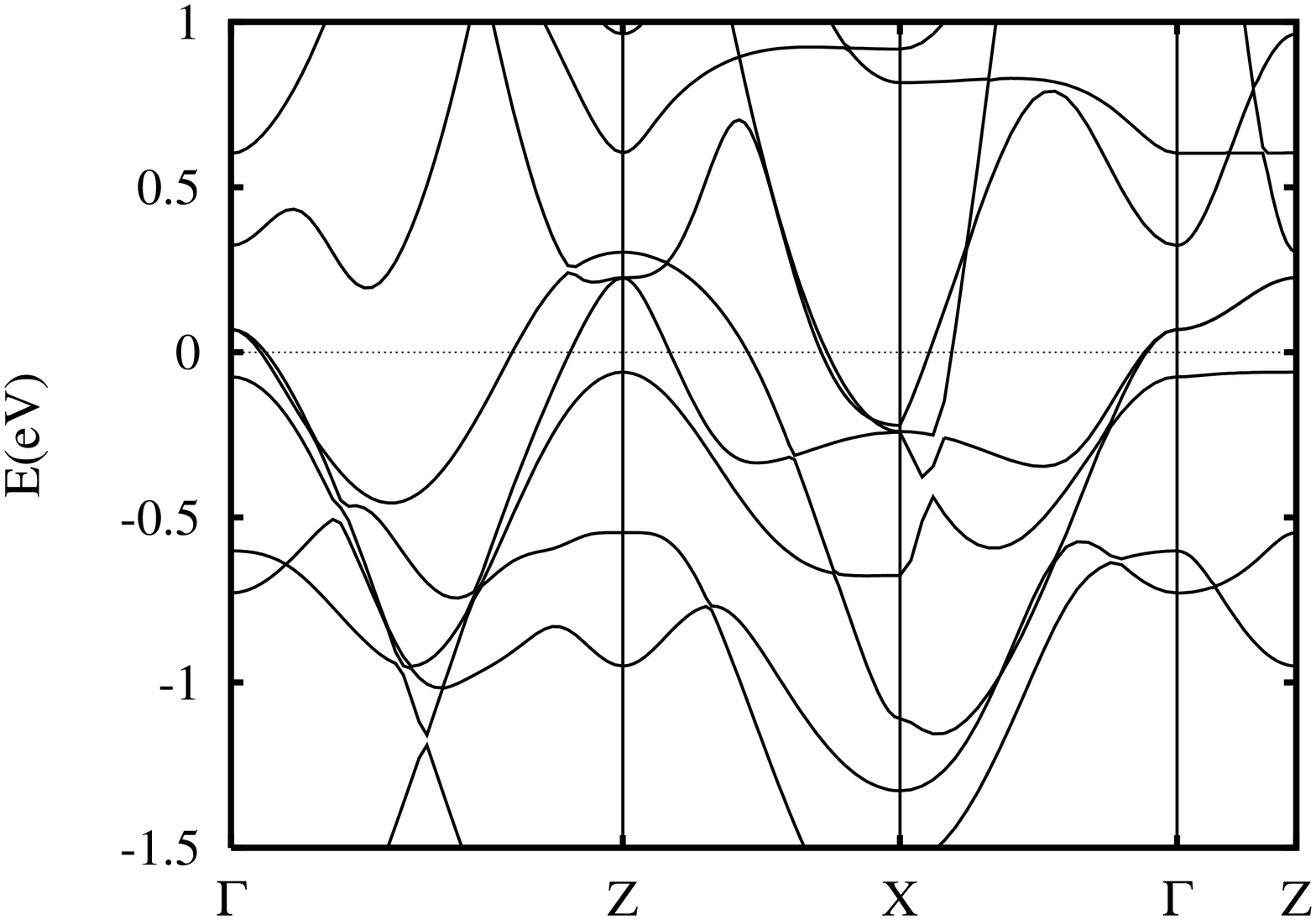}
\caption{\label{Ba-bands}
Calculated LDA band structure of BaFe$_2$As$_2$ for the LDA
value of $z_{\rm As}$. The lower panel shows a
blow-up near $E_F$, which is set to 0 eV. The short $\Gamma$-$Z$ direction
is from [0,0,0] to [0,0,1/2] while the long $\Gamma$-$Z$ is in the basal plane
and runs from [0,0,0] to [1,0,1/2] in units of the body centered tetragonal
primitive reciprocal lattice vectors.
}
\end{figure}

\begin{figure}
\includegraphics[width=3.3in,angle=0]{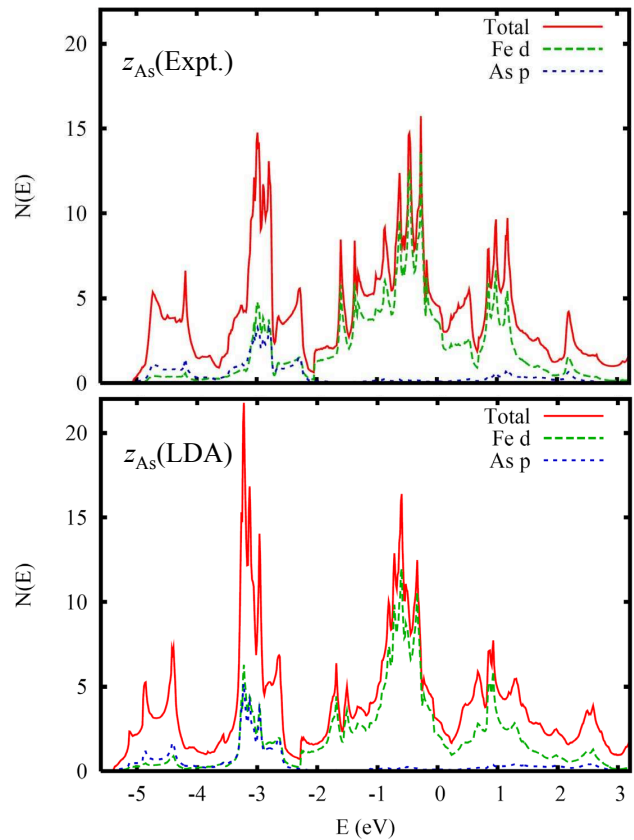}
\caption{\label{Ba-dos} (color online)
Calculated electron DOS of BaFe$_2$As$_2$ for the experimental (top)
and LDA (bottom) values of $z_{\rm As}$, plotted on a per formula unit basis.
The projections are onto the LAPW spheres. Since As has extended $p$
orbitals that extend well beyond the LAPW sphere, this yields
values that are proportional to the As contribution, but underestimate it.
}
\end{figure}

\begin{figure}
\includegraphics[height=3.3in,angle=270]{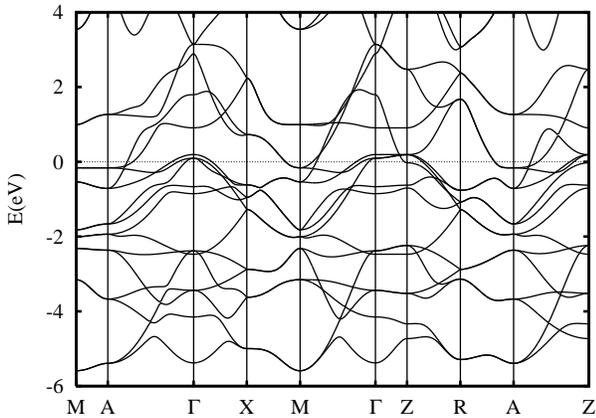}
\caption{\label{Li-bands}
Calculated LDA band structure of LiFeAs for the LDA internal coordinates.
}
\end{figure}

\begin{figure}
\includegraphics[height=3.3in,angle=270]{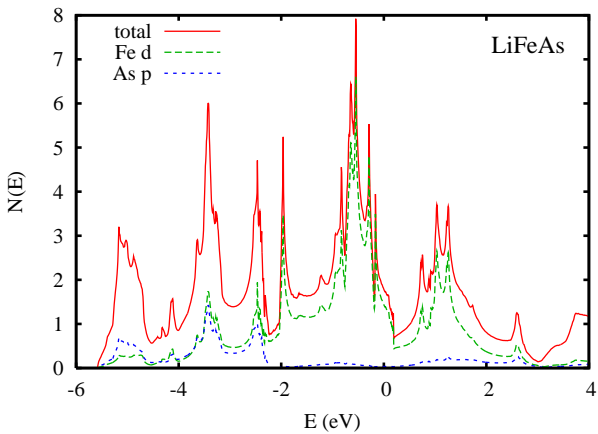}
\caption{\label{Li-dos} (color online)
Calculated electron DOS of LiFeAs for the LDA internal coordinates,
on a per formula unit basis. Note that the formula unit for this
compound contains one Fe, while that for BaFe$_2$As$_2$ contains two.
}
\end{figure}

\begin{figure}
\includegraphics[width=3.3in,angle=0]{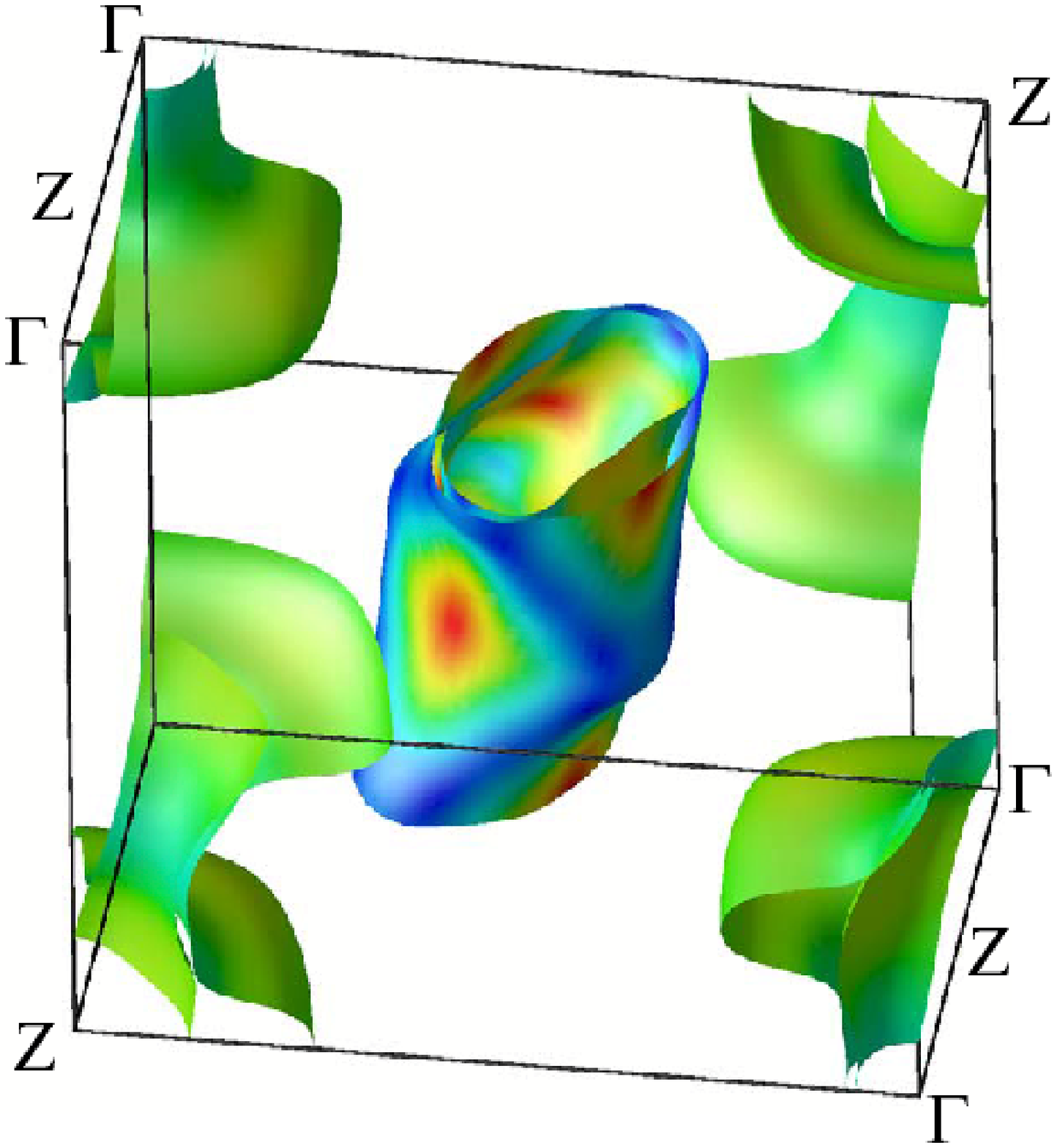}
\caption{\label{Ba-fs} (color online)
LDA Fermi surface of BaFe$_2$As$_2$ for the LDA internal coordinates,
shaded by band velocity.}
\end{figure}

\begin{figure}
\includegraphics[width=3.3in,angle=0]{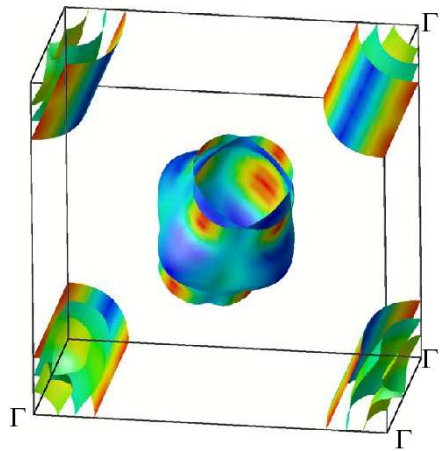}
\caption{\label{Li-fs} (color online)
LDA Fermi surface of LiFeAs for the LDA internal coordinates,
shaded by band velocity.}
\end{figure}

The calculated LDA band structures and electronic DOS for BaFe$_2$As$_2$
are given in Figs. \ref{Ba-bands} and \ref{Ba-dos}, respectively.
These were calculated using the LDA relaxed As height as given in
Table \ref{tab-struct}.
The Fermi surfaces
for the
calculated As height are depicted in Figs. \ref{Ba-fs} and \ref{Li-fs}.
As noted previously, \cite{nekrasov-basr,nekrasov-li}
They consist of electron cylinders at the zone center and hole
cylinders and other sections that depend strongly on the As height
at the zone center. 
For BaFe$_2$As$_2$ the hole Fermi surface flares out at $k_z$=1/2 ($Z$),
giving it a more three dimensional character than that of the Li compound.
This flaring out is sensitive to the As height, and in particular is
reduced as the As is moved away from the Fe plane.
In the Li compound there are three hole sections. These are two very
two dimensional hole cylinders and an inner surface that forms a capped
cylinder centered at $\Gamma$.
Calculations were also done for the experimental As height
(Fig. \ref{Ba-dos}). These
lead to a significantly higher $N(E_F)$, which will make the material more
magnetic, as has been discussed for the oxy-arsenide materials. 
\cite{mazin2,yin}
The band structure and DOS for LiFeAs are given in Figs. \ref{Li-bands}
and \ref{Li-dos}, respectively, again using the calculated internal
coordinates.
The calculated values of $N(E_F)$
are 3.06 eV$^{-1}$ per formula unit (two Fe atoms) both spins
for BaFe$_2$As$_2$ and
3.58 eV$^{-1}$ for LiFeAs on a per unit cell (two Fe atoms, two formula
units), both spins basis. For the experimental $z_{\rm As}$=0.3545
of BaFe$_2$As$_2$ we obtain $N(E_F)$ = 4.59 eV$^{-1}$ per formula unit.
These values are lower than the corresponding
value for LaFeAsO (2.62 eV$^{-1}$ per Fe both spins). \cite{singh-du}
but are still high enough to place the materials near magnetism.

For BaFe$_2$As$_2$ we find an antiferromagnetic state corresponding to the
SDW to be the most stable of the states considered. For the LDA
value of $z_{\rm As}$ we do not find any instability against either
a ferromagnetic or a checkerboard (nearest neighbor) antiferromagnetic
state in the LDA. The SDW state, which consists of lines of parallel
spin Fe atoms in the FeAs planes has a moment defined by the integral
inside the Fe LAPW sphere, radius 2.1 $a_0$, of 0.7 $\mu_B$.
As mentioned, $N(E_F)$ is larger
when the As height is raised to the reported experimental
value leading to a more magnetic state.
In this case, within the LDA we find a very weak instability of the
non-spin-polarized state to ferromagnetism (0.3 $\mu_B$/Fe, 0.01 meV/Fe).
a stronger instability for the checkerboard antiferromagnetism
(1.6 $\mu_B$/Fe, 41 meV/Fe) and the strongest instability to the SDW, which 
is the ground state (1.75 $\mu_B$/Fe, 92 meV/Fe). For the experimental value
of $z_{\rm As}$ the state with ferromagnetic $c$-axis stacking of the
Fe in the SDW state has higher energy,
and is therefore less stable than the state with antiferromagnetic
stacking, by 3 meV/Fe, while these two states are degenerate to the
precision of the calculation for the LDA As height.
Thus the lowest energy magnetic structure that we find is the SDW ordering
in the Fe layers, stacked so that Fe atoms directly above each other
in the $c$-direction are antiferromagnetically aligned.
This is in accord with recent experimental results. \cite{su}
The high sensitivity of the Fe moment to the ordering and the fact
that the SDW state is so much lower in energy than any other magnetic
state strongly imply that the magnetism is of itinerant character
with a spin density wave driven by the structure of the Fermi surface rather
than local moment physics associated with Heisenberg type exchange couplings.

These undoped electronic structures are similar to those obtained
previously and show the generic features of the calculated electronic
structures of the other FeAs materials -- 
small Fermi surfaces, with hole cylinders at the zone center
and electron cylinders at the zone corner,
high $N(E_F)$, and a strongly increasing DOS below $E_F$.
The somewhat lower values of $N(E_F)$ compared to LaFeAsO would place
BaFe$_2$As$_2$ and LiFeAs further from ferromagnetism. This combined
with the fact that ferromagnetic spin fluctuations are highly pair breaking
for singlet superconducting states may partially explain why hole doping
is effective in these non-oxide materials, but the large increase in
the DOS below $E_F$ works against this explanation taken by itself.

\section{Doped Materials: Virtual Crystal Calculations}

In this section we report virtual crystal calculations for
the two compounds. This approach differs from rigid band
in that it includes the self-consistent rearrangement of the charge
density, which is done in an average potential. This approximation
is justified since
the Ba and Li states do not contribute in any significant way to the band
structure near $E_F$ and because the scattering due to disorder
on the Ba and Li sites is expected to be weak in view of the fact
that wavefunctions of the states near $E_F$ are primarily Fe derived
and therefore separated from the Ba/Li atoms. This is supported
by the fact that, as mentioned,
we find no noticeable difference in the electronic
structure as calculated with the Li atoms of LiFeAs in sites
$2b$ and $2c$.
Importantly, in the present context, we are able to relax the
internal coordinates and in particular the As heights in the
virtual crystal approximation.

\begin{figure}
\includegraphics[height=3.3in,angle=270]{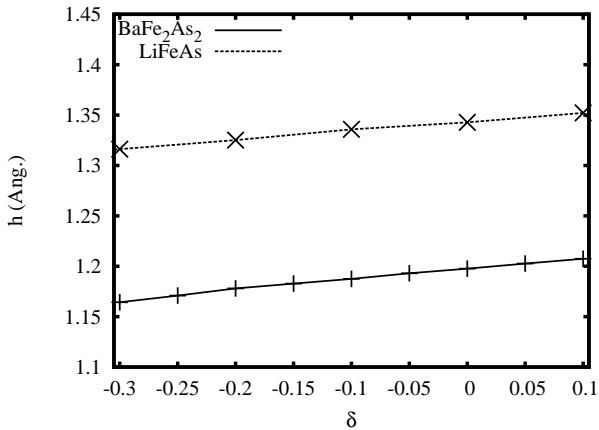}
\caption{\label{zas}
Calculated As height as a function of virtual crystal doping, $\delta$
in carriers per Fe.
Negative $\delta$ corresponds to hole doping.}
\end{figure}

The calculated As height above the Fe plane as a function of 
doping is shown in Fig. \ref{zas}. As may be seen, the As
drops towards the Fe plane as holes are introduced. This is
as might be anticipated for a primarily ionic situation, since
a more highly positively charged Fe layer will be more attractive to
As$^{3-}$ and since the effective size of transition element ions
in solids decreases as the valence increases. \cite{pauling,shannon}
The essential point is that in these materials the carrier density
is relatively low with an electronic structure
near $E_F$ formed from band edges. As a result
the DOS and Fermi surface are sensitive
to small changes, including the As position.

\begin{figure}
\includegraphics[height=3.3in,angle=270]{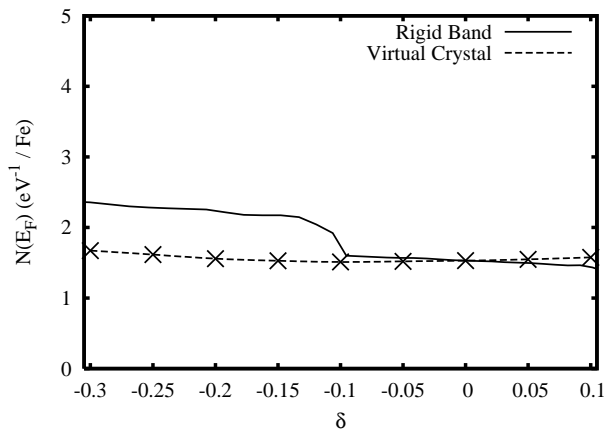}
\caption{\label{Ba-rbdos}
Rigid band DOS as compared to the virtual crystal $N(E_F)$
as a function of doping level for BaFe$_2$As$_2$.
Note that the doping levels
given in carriers per Fe with negative indicating hole doping,
and the $N(E_F)$ are given on a per Fe atom basis.}
\end{figure}

\begin{figure}
\includegraphics[height=3.3in,angle=270]{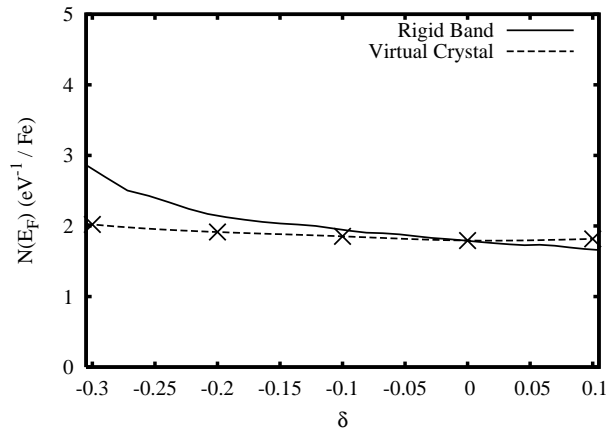}
\caption{\label{Li-rbdos}
Rigid band DOS as compared to the virtual crystal $N(E_F)$
as a function of doping level for LiFeAs.
Note that the doping levels
given in carriers per Fe with negative indicating hole doping,
and the $N(E_F)$ are given on a per Fe atom basis.}
\end{figure}

Figs. \ref{Ba-rbdos} and \ref{Li-rbdos}
show the calculated virtual crystal values of
$N(E_F)$ for various doping levels as a function of doping level.
As may be seen the behavior of the virtual crystal is very different
from what may be expected from a rigid band viewpoint. In particular,
$N(E_F)$ is found to be very nearly constant over the doping range
studied, while as mentioned, the rigid band point of view would
lead one to expect an increasing $N(E_F)$ upon hole doping.
Our result differs from that of a recent calculation of
Shein and Ivanovskii, \cite{shein} who studied a supercell
of composition Ba$_{0.5}$K$_{0.5}$Fe$_2$As$_2$, and found
a significant increase in $N(E_F)$ relative to the undoped compound.
We ascribe this difference to the fact that they used the fixed
experimental As position for the undoped compound and did not relax
the internal coordinates.
In any case, we find that contrary to the rigid band picture hole doping does
not bring the system closer to Stoner
ferromagnetism, which as mentioned would be detrimental to superconductivity.
Thus the main effects on the electronic structure are more subtle.

\begin{figure}
\includegraphics[width=3.3in,angle=0]{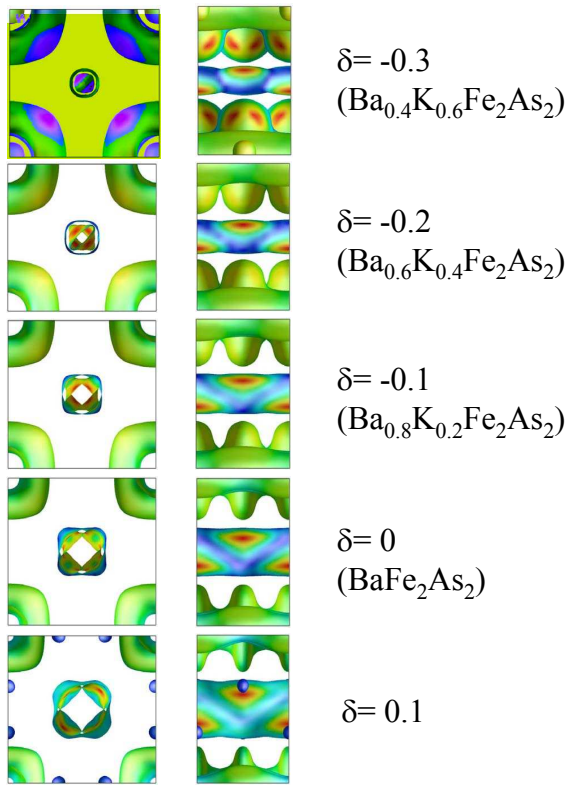}
\caption{\label{Ba-fsdope} (color online)
Virtual crystal Fermi surface of BaFe$_2$As$_2$ as a function of
doping level $\delta$ using the relaxed As position for each value
of $\delta$. The shading is by velocity.
The left panels show a view along the $k_z$ directions,
while the right panels are along $k_x$.
}
\end{figure}

\begin{figure}
\includegraphics[width=3.3in,angle=0]{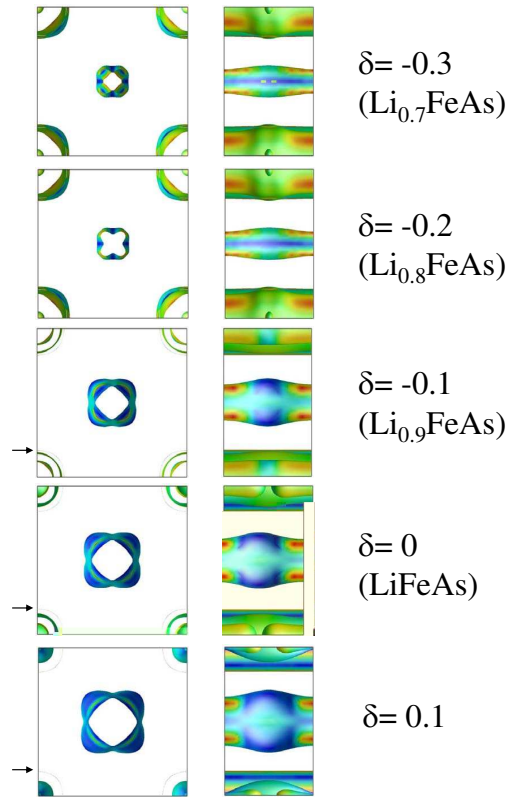}
\caption{\label{Li-fsdope} (color online)
Virtual crystal Fermi surface of LiFeAs as a function of
doping level $\delta$ using the relaxed atomic positions for each value
of $\delta$. The shading is by velocity.
The left panels show a view along the $k_z$ directions,
while the right panels are along $k_x$.
The outermost hole cylinder is very two dimensional and therefore is
difficult to see in the view along $k_z$. Accordingly we indicate its
position by an arrow.
}
\end{figure}

Figs. \ref{Ba-fsdope} and \ref{Li-fsdope}
show the evolution of the non-spin-polarized
virtual crystal Fermi surface for the two compounds.
As may be seen the basic structure of the Fermi surface
with hole sections around the zone center and electron sections at
the zone corner is maintained for all the doping levels shown.
The difference between the volume of the hole and electron sections
is required to correspond to the doping level by the Luttinger theorem.
Also it may be seen that the size of the electron sections changes
at least
as much as that of the hole sections (note that in Fig. \ref{Ba-fsdope}
the top view emphasizes the flaring out of the hole Fermi surface
at $k_z$=0.5, while the main cylindrical sections are smaller).
This is contrary to what
might be anticipated considering that the electron bands
are more dispersive than the hole bands and therefore
should change less and again shows the importance
of non-rigid band effects.

The condition for an SDW instability of the paramagnetic state
is that the real part of the susceptibility, $\chi({\bf q})$
should diverge at the nesting vector.
This function is given in terms of an integral over the Fermi surface
for the bare Lindhard susceptibility, $\chi_0({\bf q})$,
with an enhancement factor that emphasizes peaks, much like the Stoner
enhancement of the Pauli susceptibility for the ferromagnetic case.
The interband
Lindhard susceptibility for an electron
cylinder at the zone corner and a same sized hole cylinder at the zone
center is a function peaked at the nesting vector, while if the cylinders
differ in radius by $\delta q$, e.g. due to doping or extra hole sections,
the function will have a flat cylindrical plateau centered at the former
nesting vector with diameter $2\delta q$. Thus as the cylinders become
mismatched in size the maximum value of $\chi_0({\bf q})$
(the value at the nesting vector) will decrease, while a region of high
constant $\chi_0({\bf q})$ will develop within $\delta q$ of the nesting
vector.

We calculated the Lindhard function, setting all matrix elements to be
equal (i.e. constant matrix element approximation yielding
arbitrary units).
Although the Fermi surfaces are not perfect cylinders and there
are multiple surfaces of different size, we do in fact find this
behavior for LiFeAs. It is also the case for electron doping in
BaFe$_2$As$_2$ and to a lesser degree for hole doped BaFe$_2$As$_2$,
where the peak in $\chi_0({\bf q})$ becomes broader and smaller but there
is no clear plateau region. For both LiFeAs and BaFe$_2$As$_2$
we find a decrease in the peak height of the Lindhard function
as we dope with holes. This explains why the SDW is destroyed by hole
doping in these materials.

\section{discussion}

There has been much recent discussion of the relationship
between the SDW instability and superconductivity.
Within an Eliashberg formalism involving spin fluctuations what enters
the calculation of the pairing function is a double integral over
the Fermi surface (like that in the Lindhard function) with a kernel
containing the pairing interaction, $V({\bf q}) \sim \chi({\bf q})$.
Thus in the case discussed above, the
interband superconducting pairing between the electron and
hole Fermi surfaces will depend on an integral over a range of
${\bf q}$ of size comparable to the Fermi surface size, while
the SDW instability will depend on the peak value of $\chi({\bf q})$
(in a real case
spin fluctuations away from the SDW ordering vector will compete
with the SDW so even for a given maximum value of the bare $\chi_0({\bf q})$
a system with a narrower peak will order first).
In any case,
if spin fluctuations associated with the SDW
are responsible for pairing, one expects superconductivity to compete with the
SDW and to occur on both sides of the region where the SDW is stable.
What our results show is that the properties are much more symmetric
between hole and electron doping than would be expected from a rigid band
picture.

\acknowledgements

We are grateful for helpful discussions with I.I. Mazin,
M.H. Du, D.G. Mandrus and B.C. Sales.
This work was supported by the Department of Energy, Division of
Materials Sciences and Engineering.

\end{document}